\documentclass[%
 reprint,
 amsmath,amssymb,
 aps,
 prb,
 floatfix,
]{revtex4-2}

\usepackage{graphicx}
\usepackage{dcolumn}
\usepackage{bm}
\usepackage{array}
\usepackage{tabularx}
\usepackage{tikz}
\usetikzlibrary{decorations.pathmorphing,decorations.markings,arrows.meta,positioning,calc,shapes.geometric}
\usepackage[colorlinks=true,linkcolor=blue,citecolor=blue,urlcolor=blue]{hyperref}

\newcommand{\ii}{\mathrm{i}}
\newcommand{\dd}{\mathrm{d}}

\newcommand{\Ree}{\operatorname{Re}}
\newcommand{\Imm}{\operatorname{Im}}

\tikzset{
  >={Stealth[scale=0.9]},
  electron/.style={thick, postaction={decorate},
    decoration={markings, mark=at position 0.55 with {\arrow{>}}}},
  phonon/.style={thick, decorate,
    decoration={snake, amplitude=0.7mm, segment length=1.8mm,
                pre length=0.4mm, post length=0.4mm}},
  dphi/.style={thick, double, double distance=1.3pt, decorate,
    decoration={snake, amplitude=0.85mm, segment length=2.0mm,
                pre length=0.4mm, post length=0.4mm}},
  vc/.style={thick},
  blob/.style={circle, draw, fill=gray!30, thick,
               inner sep=0.5pt, minimum size=6mm, font=\small},
  sigmablob/.style={circle, draw, fill=gray!30, thick,
                    inner sep=0pt, minimum size=7mm},
  dot/.style={circle, fill=black, inner sep=0pt, minimum size=2pt},
}

\begin{document}

\title{Microscopic Theory of Acoustic Phonon Scattering by Charge-Density-Wave Fluctuations}

\author{Han Huang}
 \email{hh697@cornell.edu}
\affiliation{Sibley School of Mechanical and Aerospace Engineering, Cornell University, Ithaca, New York 14853, USA}

\date{\today}

\begin{abstract}
Charge-density-wave (CDW) order in correlated metals originates in a peaked electronic susceptibility at a finite wavevector $\mathbf Q_0$, set either by Fermi-surface features (nesting or saddle-point singularities) or by momentum-resolved electron-phonon coupling, or by a combination of the two.  CDW precursor fluctuations can attenuate heat-carrying acoustic phonons even when long-range order is absent.  We develop a Green's-function theory in which a damped-harmonic-oscillator propagator for a hybrid CDW--lattice soft mode at the ordering wavevector $\mathbf Q_0$ and a strain--intensity vertex obtained from an electron loop combine to give the acoustic phonon self-energy.  The theory identifies two scattering channels: a local-intensity channel, controlled by a retarded composite CDW response and giving a narrow critical contribution when the CDW correlation length is large, and a texture (gradient) channel, which couples acoustic strain to spatial variations of the CDW envelope and, in a frozen-texture limit, reduces to a phenomenological form set by the measured diffraction peak weight and width.  The same propagator fixes the lattice projection of a hybrid CDW--phonon soft pole measured by inelastic X-ray scattering, with an underdamped-to-overdamped crossover controlled by the distance to the CDW instability and a mass-tracking identity for the slow overdamped relaxation rate.  The framework unifies diffraction, soft-mode spectroscopy, and thermal transport and applies broadly across CDW materials, including the transition-metal dichalcogenides, rare-earth tritellurides, kagome CDW compounds, and the cuprate fluctuating charge-order regime; we illustrate it by direct comparison with experimental IXS phonon softening and anomalous thermal transport in 2H-TaSe$_2$ at elevated temperatures.
\end{abstract}

\maketitle

\section{Introduction}

Fluctuations of an incipient order parameter dominate the low-energy physics of a metal when the electron system approaches an instability but has not yet broken the associated symmetry. In nearly ferromagnetic metals, soft particle-hole spin fluctuations (paramagnons) mediate an effective interaction between conduction electrons and leave distinctive signatures in thermodynamics, transport, and superconducting pairing~\cite{BerkSchrieffer,DoniachEngelsberg,Moriya}. In metals sitting close to a charge-density-wave (CDW) instability, an analogous bosonic mode develops: a soft, long-lived lattice distortion with finite correlation length that emerges well above the ordering temperature and behaves as an overdamped propagating degree of freedom in its own right~\cite{Gruner1988,Monceau2012}. The present work develops a microscopic theory of acoustic phonon scattering by such CDW fluctuations, casting the CDW precursor as the charge-sector analog of the near-ferromagnetic mode.

The paramagnon concept was originally invoked to explain the suppression of $s$-wave superconductivity and the enhanced low-temperature specific heat and resistivity of Pd and Pd-based alloys~\cite{BerkSchrieffer,DoniachEngelsberg}. In a nearly ferromagnetic Fermi liquid, the exchange interaction drives the uniform spin susceptibility close to a Stoner divergence, so that the particle-hole spin-fluctuation propagator acquires a large, weakly damped pole. Scattering of quasiparticles by these fluctuations yields enhancements to the nuclear spin-lattice relaxation rate~\cite{BealMonod1968}, a $T\ln T$ correction to the specific heat, and a $T^{2}$ resistivity with an anomalously large coefficient, all controlled by the proximity to the Stoner instability~\cite{DoniachEngelsberg,Moriya}. The same overdamped spin-fluctuation mode can mediate unconventional, non-$s$-wave pairing that avoids the short-range repulsion responsible for quenching phonon-driven superconductivity. This reasoning was initiated for the $p$-wave state of $^{3}$He~\cite{FayLayzer1968,AndersonBrinkman1973}, extended to the $d$-wave pairing of heavy-fermion compounds~\cite{Miyake1986}, and now stands as a working framework for the cuprates, iron pnictides, and other unconventional superconductors~\cite{Monthoux2007,Scalapino2012}.

A parallel story unfolds in the charge sector. In a metal with partially nested Fermi surface or with strong $\mathbf{q}$-dependent electron-phonon coupling, a phonon branch softens at a wavevector $\mathbf{Q}_{0}$ on cooling and eventually condenses into a static periodic lattice distortion with associated charge modulation at the CDW transition temperature $T_{\mathrm{CDW}}$~\cite{Gruner1988,Monceau2012}. Above $T_{\mathrm{CDW}}$, the phonon at $\mathbf{Q}_{0}$ is not simply a harmonic excitation of the high-symmetry lattice but an overdamped, finite-correlation-length fluctuation of the would-be broken symmetry: the charge-sector counterpart of the paramagnon. The canonical transition-metal dichalcogenides 2H-TaSe$_{2}$ and 2H-NbSe$_{2}$ exhibit this scenario in a particularly clean form~\cite{MoncAxeDiSalvo1977}. The simple nesting picture has given way to momentum-resolved electron-phonon coupling~\cite{JohannesMazin2008,Rossnagel2011} and, in strongly anharmonic cases, to double-well Ising scenarios~\cite{Gorkov2012}. The same phenomenology recurs across a wide range of platforms: the rare-earth tritellurides $R$Te$_{3}$~\cite{Brouet2008}, the fluctuating charge order of underdoped cuprates~\cite{Ghiringhelli2012,Chang2012}, and the recently discovered kagome superconductor CsV$_{3}$Sb$_{5}$~\cite{Ortiz2020}.

2H-TaSe$_{2}$ occupies a distinguished place in this family: its incommensurate CDW transition at $T_{\mathrm{CDW}}\approx 122$\,K is preceded by an unusually extended fluctuation region in which signatures of incipient order persist up to nearly twice $T_{\mathrm{CDW}}$. Recent high-resolution inelastic x-ray scattering (IXS) has shown that the in-plane acoustic branch softens completely to zero energy at $\mathbf{Q}_{0}$ already tens of kelvin above $T_{\mathrm{CDW}}$, while remaining strongly dispersive and long-lived at other wavevectors~\cite{Weber2023}. Complementary transient thermal grating (TTG) measurements probe the low-wavevector acoustic thermal transport in the same crystal across the same temperature window~\cite{Huang2026}. The TTG technique resolves acoustic phonon dynamics at gigahertz frequencies and micron wavelengths, $|\mathbf{q}|\ll|\mathbf{Q}_{0}|$, and is therefore sensitive to the long-wavelength tail of any scattering process whose rate is set by the CDW fluctuation spectrum~\cite{Maznev2011}. Those measurements revealed a pronounced, non-monotonic temperature dependence of the acoustic thermal diffusivity that tracks the precursor region identified by IXS, providing direct evidence that CDW fluctuations scatter acoustic phonons with anomalous temperature dependence.

Despite this accumulated experimental consensus, a quantitative microscopic description of acoustic phonon scattering by CDW fluctuations, one that would allow the IXS and TTG observations to be tied together on a common footing, has remained conspicuously underdeveloped. The paramagnon literature provides a ready template. The quasiparticle self-energy is obtained by convolving a fermion propagator with an overdamped bosonic susceptibility; the resulting transport coefficients and effective mass acquire their characteristic non-Fermi-liquid temperature dependences from the boson's correlation length $\xi(T)$~\cite{Moriya,Scalapino2012}. The CDW problem has an almost identical kinematic structure, with acoustic phonons playing the role of the scattered quasiparticles and the soft CDW mode playing the role of the boson. Because the probe couples to strain rather than spin, the vertex is fixed by the elastic and electron-phonon response of the material, quantities accessible from density-functional perturbation theory~\cite{Baroni2001,Giustino2017}, and the only free inputs are the temperature-dependent CDW correlation length and linewidth, which can be taken directly from IXS. This observation is the central premise of the present work.

In this paper we construct the microscopic theory of acoustic phonon scattering by CDW fluctuations and evaluate it for 2H-TaSe$_{2}$.  We begin by introducing an effective action for a hybrid CDW--lattice soft mode near $\mathbf Q_0$, whose damped-harmonic-oscillator propagator tracks the approach to the CDW instability through a temperature-dependent mass parameter $r(T)$ and exhibits an underdamped-to-overdamped crossover whose slow relaxation rate obeys a mass-tracking identity.  From the electron triangle we derive a strain--intensity vertex with a local coupling to $|\Phi|^{2}$ and a momentum-gradient coupling to the gradient bilinear of $\Phi$.  We then compute the acoustic phonon self-energy and lifetime in the long-wavelength limit probed by TTG, separating the contribution into a local-intensity channel set by a retarded composite CDW susceptibility and giving a narrow critical contribution when the CDW correlation length is large, and a texture (gradient) channel dominated by spatial variations of the CDW envelope and reducing in a frozen-texture limit to the phenomenological form fixed by the diffraction peak weight and width.  The same propagator also determines the lattice projection of a hybrid CDW--phonon soft pole resolved by inelastic X-ray scattering.  The framework can be transferred to the broader family of CDW materials introduced above.

\section{Theory}

\subsection{Hybrid CDW--lattice soft mode}

At the phenomenological level we describe a CDW near wavevector $\mathbf Q_0$ by a slowly varying complex order-parameter field $\psi(\mathbf r)$: its magnitude $|\psi|$ sets the local amplitude of the charge modulation and its phase $\arg\psi$ sets the local registry, so the full electronic density reads
\begin{equation}
\delta\rho(\mathbf r)\simeq \psi(\mathbf r)e^{i\mathbf Q_0\cdot\mathbf r}
+\psi^*(\mathbf r)e^{-i\mathbf Q_0\cdot\mathbf r}.
\label{eq:cdw_envelope}
\end{equation}
The field $\psi$ is the coarse-grained CDW collective mode whose dynamics, correlation length, and coupling to strain we will compute microscopically below.

The low-energy theory begins with an attractive charge-channel interaction
\begin{align}
S_{\rho}=&-\frac12 T\sum_{\Omega_n}\int\frac{\dd^dq}{(2\pi)^d}\,
g_c(\mathbf q,\ii\Omega_n)
\nonumber\\
&\times \rho(\mathbf q,\ii\Omega_n)\rho(-\mathbf q,-\ii\Omega_n),
\label{eq:Srho}
\end{align}
Here $\Omega_n=2\pi n T$ is a bosonic Matsubara frequency, $d$ is the spatial dimension, $\rho(\mathbf q,\ii\Omega_n)$ is the electron density at wavevector $\mathbf q$ and frequency $\Omega_n$, and $g_c(\mathbf q,\ii\Omega_n)>0$ is the effective attraction in the CDW channel.  Retarded optical phonons, Coulomb-renormalized electron-phonon coupling, and electronic vertex structure can all be included in this single kernel; no separate signed interaction is introduced.

Decoupling the four-fermion CDW channel near $\mathbf Q_0$ with a Hubbard--Stratonovich field $\Phi_P$ conjugate to $\rho_{\mathbf Q_0+\mathbf p}$ (where $P\equiv(\mathbf p,\ii\nu_n)$ is a shorthand for the small momentum measured from $\mathbf Q_0$ together with the bosonic Matsubara frequency) and integrating out the fermions yields the effective action
\begin{equation}
S_{\rm eff}[\Phi]=\sum_P g_c^{-1}(\mathbf Q_0+\mathbf p,\ii\nu_n)|\Phi_P|^2-\mathrm{Tr}\ln\!\left(\hat G_0^{-1}-\hat\Sigma_\Phi\right),
\label{eq:Seff_HS}
\end{equation}
where $\hat G_0$ is the bare electron Matsubara Green's function and $\hat\Sigma_\Phi$ is the self-energy operator linear in $\Phi$ generated by the decoupling.  The HS field $\Phi$ enters microscopically only as an auxiliary bosonic variable in the CDW channel, but once the fermions have been integrated out, Eq.~\eqref{eq:Seff_HS} is precisely the action of the fluctuating CDW collective mode.  It is therefore proportional, up to a normalization constant absorbed into $g_c$ and the vertices below, to the coarse-grained order parameter $\psi$ of Eq.~\eqref{eq:cdw_envelope}; from here on we identify the two and continue to use the symbol $\Phi$ for both roles.  Expanding the trace logarithm to Gaussian order in $\Phi$ gives the bare CDW propagator
\begin{equation}
D_\Phi^{-1}(\mathbf q,\ii\nu_n)=g_c^{-1}(\mathbf q,\ii\nu_n)-\Pi(\mathbf q,\ii\nu_n).
\label{eq:Dphi_inv}
\end{equation}
For a multiband metal,
\begin{align}
\Pi(\mathbf q,\ii\nu_n)=&-T\sum_{\omega_n}\int\frac{\dd^dk}{(2\pi)^d}
\mathrm{Tr}\left[
M_{\mathbf q}G_0(\mathbf k+\mathbf q,\ii\omega_n+\ii\nu_n)
\right.
\nonumber\\
&\hspace{2.2cm}\left.
\times M_{\mathbf q}^\dagger G_0(\mathbf k,\ii\omega_n)
\right],
\label{eq:Pi_multiband}
\end{align}
where $G_0$ is the bare electron Matsubara Green's function and $M_{\mathbf q}$ is the CDW form factor, the band-resolved matrix element that converts the electron density into the CDW order parameter at wavevector $\mathbf q$.  In a one-band model the fermionic Matsubara sum gives
\begin{equation}
\Pi(\mathbf q,\ii\nu_n)=
-\int\frac{\dd^dk}{(2\pi)^d}
\frac{f(\varepsilon_{\mathbf k})-f(\varepsilon_{\mathbf k+\mathbf q})}
{\ii\nu_n+\varepsilon_{\mathbf k}-\varepsilon_{\mathbf k+\mathbf q}} .
\label{eq:Pi_sum}
\end{equation}
After analytic continuation, this is the usual Lindhard response with the chosen sign convention.

In an IXS experiment the measured object is not a purely electronic density propagator.  The electronic CDW coordinate and the lattice displacement coordinate near $\mathbf Q_0$ are hybridized~\cite{Weber2023}.  Keeping both degrees of freedom as independent fields, stacked in a $(\rho,u)$ basis, the retarded inverse propagator takes the $2\times 2$ form
\begin{equation}
\mathcal G_{\rho u}^{-1}=
\begin{pmatrix}
\chi_\rho^{-1} & -g_{\rho u}\\
-g_{\rho u} & D_{u,0}^{-1}
\end{pmatrix}.
\label{eq:hybrid_matrix}
\end{equation}
Here $\chi_\rho(\mathbf Q_0,\omega)$ is the retarded bare electronic density susceptibility in the CDW channel, obtained by analytically continuing the Matsubara Lindhard bubble $\Pi(\mathbf Q_0,\ii\nu_n)$ of Eq.~\eqref{eq:Pi_multiband} (or its multiband generalization) to real frequency, $\ii\nu_n\to\omega+\ii 0^+$; $D_{u,0}(\mathbf Q_0,\omega)$ is the bare acoustic-phonon propagator, and $g_{\rho u}$ is the linear charge-lattice coupling; all three are complex scalars obtained from the corresponding microscopic responses.  Inverting the matrix gives the diagonal lattice--lattice and charge--charge response functions
\begin{align}
G_{uu}^R&=\frac{\chi_\rho^{-1}}
{\chi_\rho^{-1}D_{u,0}^{-1}-g_{\rho u}^2},
\nonumber\\
G_{\rho\rho}^R&=\frac{D_{u,0}^{-1}}
{\chi_\rho^{-1}D_{u,0}^{-1}-g_{\rho u}^2}.
\label{eq:shared_poles}
\end{align}
These two responses share the same pole but carry different spectral weights.  The lattice-lattice element $G_{uu}^R$, which is what ordinary IXS measures, is read off the inverted $2\times 2$ hybrid matrix in the $u$-$u$ slot; the scalar $D_\Phi$ used below denotes the reduced low-energy propagator for the shared hybrid soft pole.

Expanding Eq.~\eqref{eq:Dphi_inv} around the soft wavevector, $\mathbf q=\mathbf Q_0+\mathbf p$, gives
\begin{equation}
D_\Phi^{-1}(\mathbf Q_0+\mathbf p,\ii\nu_n)
=r+c p^2+\chi_\omega\nu_n^2+\eta|\nu_n|+\cdots .
\label{eq:Dphi_expanded}
\end{equation}
The four coefficients are real constants, each capturing a distinct feature of the soft mode: $r$ is the static CDW mass (inverse uniform susceptibility at $\mathbf Q_0$), $c$ is the spatial stiffness, $\chi_\omega$ is the inertial coefficient, and $\eta$ is the Landau-damping rate, all read off from the Taylor expansion of $g_c^{-1}-\Pi$ about $(\mathbf Q_0,0)$.  The soft-mode mass (inverse static CDW susceptibility) is
\begin{equation}
r(T)=g_c^{-1}(\mathbf Q_0,0)-\Pi(\mathbf Q_0,0;T).
\label{eq:rdef}
\end{equation}
The absence of a term linear in $\mathbf p$ follows from choosing $\mathbf Q_0$ at the minimum of the static inverse propagator.  The $\nu_n^2$ term gives inertial soft-mode dynamics; the $|\nu_n|$ term gives overdamped relaxation.

At zero frequency,
\begin{equation}
D_\Phi^R(\mathbf Q_0+\mathbf p,0)=\frac{1}{r+c p^2}
=\frac{1/c}{p^2+\xi^{-2}},
\label{eq:xi_def}
\end{equation}
so the Gaussian CDW correlation length is
\begin{equation}
\xi=\sqrt{\frac{c}{r}},\qquad r=c\xi^{-2}.
\label{eq:xi_r}
\end{equation}
Here $\xi^{-1}$ is the momentum width of the static fluctuation peak around $\mathbf Q_0$.  The continuum cutoff is denoted $\Lambda$.  The condition $\Lambda\xi\gg1$ defines the long-correlation regime in which infrared scaling is meaningful; $\Lambda\xi\ll1$ is a high-mass cutoff expansion.

After analytic continuation,
\begin{equation}
D_\Phi^{-1,R}(\mathbf Q_0+\mathbf p,\omega)
=r+c p^2-\chi_\omega\omega^2-\ii\eta\omega+\cdots .
\label{eq:Dphi_ret}
\end{equation}
If the mode is overdamped,
\begin{equation}
D_\Phi^R(\mathbf Q_0+\mathbf p,\omega)\simeq
\frac{1}{r+c p^2-\ii\eta\omega},
\label{eq:D_overdamped}
\end{equation}
and its softest relaxation time is
\begin{equation}
\tau_{\rm rel}=\frac{\eta}{r}.
\label{eq:taurel}
\end{equation}
The low-frequency CDW hydrodynamic limit used below requires both $K\xi\ll1$ and $\Omega\tau_{\rm rel}\ll1$, where $\mathbf K$ and $\Omega$ are the acoustic phonon wavevector and frequency.  If the mode remains underdamped, the same pole can be written as
\begin{align}
D_\Phi^{-1,R}&\simeq
\chi_\omega\left[\widetilde\omega_q^2-\omega^2-2\ii\Gamma_q\omega\right],
\nonumber\\
\widetilde\omega_q^2&=\frac{r+c p^2}{\chi_\omega},
\qquad
\Gamma_q=\frac{\eta}{2\chi_\omega}.
\label{eq:DHO}
\end{align}

\subsection{Strain vertex and acoustic phonon self-energy}

Acoustic phonons are described by the atomic displacement field $u_i(\mathbf x)$, and they enter the electronic problem only through its symmetrized gradient, the strain tensor,
\begin{equation}
\epsilon_{ij}=\frac12(\partial_i u_j+\partial_j u_i).
\label{eq:strain}
\end{equation}
Here $\widetilde\omega_q^2=(r+c\mathbf p^2)/\chi_\omega$ is the squared natural frequency of the soft hybrid CDW--lattice mode, with $r=r(T)$ measuring the distance to the CDW instability through Eq.~\eqref{eq:rdef} and $c$ the gradient stiffness, while $2\Gamma_q=\eta/\chi_\omega$ is its damping rate, built from the relaxation coefficient $\eta$ and the inertial response $\chi_\omega$.  The mode is underdamped when $\widetilde\omega_q>\Gamma_q$ and overdamped when $\widetilde\omega_q<\Gamma_q$; the crossover is controlled by $r(T)$, since $\widetilde\omega_q^2\propto r$ at $\mathbf p=0$, so that approaching the CDW instability necessarily drives the soft mode through the underdamped--to--overdamped crossover.

The electron-strain coupling is written as
\begin{align}
S_{e-\epsilon}=&T\sum_{\omega_n,\Omega_m}
\int\frac{\dd^dk}{(2\pi)^d}\frac{\dd^dK}{(2\pi)^d}\,
\epsilon_{ij}(\mathbf K,\ii\Omega_m)\gamma_{ij}(\mathbf k;\mathbf K)
\nonumber\\
&\times
\bar c(\mathbf k+\mathbf K,\ii\omega_n+\ii\Omega_m)
c(\mathbf k,\ii\omega_n).
\label{eq:Seps}
\end{align}
Here $c,\bar c$ are the conduction-electron fields and $\gamma_{ij}(\mathbf k;\mathbf K)$ is the bare electron-strain matrix element, which plays the same role for strain that $M_{\mathbf q}$ plays for the CDW order parameter.  Expanding the fermion trace to second order in $\Phi$ and first order in strain gives
\begin{align}
S_{u\Phi}=&T\sum_{\nu_n,\Omega_m}
\int\frac{\dd^dp}{(2\pi)^d}\frac{\dd^dK}{(2\pi)^d}\,
\epsilon_{ij}(\mathbf K,\ii\Omega_m)
\nonumber\\
&\times
\Gamma_{ij}(\mathbf p,\ii\nu_n;\mathbf K,\ii\Omega_m)
\Phi^*(\mathbf Q_0+\mathbf p+\mathbf K,\ii\nu_n+\ii\Omega_m)
\nonumber\\
&\times
\Phi(\mathbf Q_0+\mathbf p,\ii\nu_n).
\label{eq:SuPhi}
\end{align}
The coefficient $\Gamma_{ij}(\mathbf p,\ii\nu_n;\mathbf K,\ii\Omega_m)$ is a fermion-loop integral with three internal electron propagators: two of them carry $\Phi$ legs and one carries the strain leg, giving the diagram the shape of a triangle.  Only a single internal electron momentum-frequency variable is left after the three external bosonic legs are fixed; a convenient routing labels the three internal propagators by
\begin{align}
k_1&=(\mathbf k,\ii\omega_n),\nonumber\\
k_2&=(\mathbf k-\mathbf Q_0-\mathbf p-\mathbf K,\ii\omega_n-\ii\nu_n-\ii\Omega_m),\nonumber\\
k_3&=(\mathbf k-\mathbf K,\ii\omega_n-\ii\Omega_m).
\label{eq:routing}
\end{align}
These are not additional integration variables; they are the three electron-line arguments after the single loop variable has been chosen.  With this routing,
\begin{align}
\Gamma_{ij}=&T\sum_{\omega_n}\int\frac{\dd^dk}{(2\pi)^d}
\mathrm{Tr}\Big[
\gamma_{ij}(k_1,k_3)G_0(k_1)
\nonumber\\
&\hspace{1.0cm}\times
M^\dagger G_0(k_2)M G_0(k_3)
\Big]
+\mathrm{perms.}+{\rm H.c.}
\label{eq:Gamma}
\end{align}
Each of the three internal propagators $G_0(k_j)$ in Eq.~\eqref{eq:Gamma} is a free electron line, which on the imaginary-frequency axis has a single pole at $\ii\omega_n=z_j$, where
\begin{equation}
z_j\equiv \varepsilon(\mathbf k_j)-\mu+(\text{external bosonic shift on leg }j)
\label{eq:zj_def}
\end{equation}
is the band energy on that leg measured from the chemical potential, supplemented by whichever external frequencies the routing of Eq.~\eqref{eq:routing} assigned to it.  These three poles $z_1,z_2,z_3$ are the only singularities of the integrand in the complex $\ii\omega_n$ plane.  The loop frequency $\ii\omega_n$ enters $\Gamma_{ij}$ of Eq.~\eqref{eq:Gamma} only through the three electron denominators, so it can be eliminated in closed form by the standard triangle Matsubara sum,
\begin{equation}
T\sum_{\omega_n}\prod_{j=1}^3\frac{1}{\ii\omega_n-z_j}
=
\sum_{j=1}^3
\frac{f(z_j)}{\prod_{\ell\neq j}(z_j-z_\ell)},
\label{eq:triangle_sum}
\end{equation}
with $f(z)$ the Fermi distribution.  Carrying out this sum collapses $\Gamma_{ij}$ to a single band-momentum integral whose integrand combines the band energies on the three routed lines, their Fermi factors, and the bare strain matrix element $\gamma_{ij}$ introduced below Eq.~\eqref{eq:Seps}. 

At long wavelengths the triangle of Eq.~\eqref{eq:Gamma} can be expanded in gradients of $\Phi$, giving a local effective action in which strain couples only to two scalar operators built from the order parameter,
\begin{align}
S_{u\Phi}=&\int\dd^dx\,\epsilon_{ij}(\mathbf x)
\left[\lambda_0^{ij}|\Phi|^2
+\lambda_1^{ij;ab}\mathcal T_{ab}
+\cdots\right],
\label{eq:real_vertex}\\
\mathcal T_{ab}=&\,
\Ree\left[(\partial_a\Phi)^*(\partial_b\Phi)\right].
\label{eq:Gab_def}
\end{align}
The real coupling tensors $\lambda_0^{ij}$ and $\lambda_1^{ij;ab}$ are material-specific constants, fixed by evaluating the triangle at zero external momentum and expanding to zeroth and second order, respectively, in the small momentum $\mathbf p$ carried by $\Phi$; they have the same symmetries as the corresponding elastic moduli.  The first term couples strain to the local CDW intensity $|\Phi|^2$.  The second couples strain to the gradient bilinear $\mathcal T_{ab}$, a real symmetric tensor built from the spatial variation of $\Phi$.  For $\Phi=Ae^{\ii\theta}$,
\begin{equation}
\mathcal T_{ab}=
\partial_aA\,\partial_bA+A^2\partial_a\theta\,\partial_b\theta .
\label{eq:grad_decomp}
\end{equation}
The gradient coupling $\mathcal T_{ab}$ therefore probes both the amplitude texture $\partial_a A\,\partial_b A$ and the phase texture $A^2\partial_a\theta\,\partial_b\theta$ of the CDW, on an equal footing.

Two such vertices give the strain self-energy
\begin{align}
\Sigma_{\epsilon}^{ij,kl}(\mathbf K,\ii\Omega_m)=&
T\sum_{\nu_n}\int\frac{\dd^dp}{(2\pi)^d}
\Gamma_{ij}(\mathbf p,\ii\nu_n;\mathbf K,\ii\Omega_m)
\nonumber\\
&\times D_\Phi(\mathbf Q_0+\mathbf p,\ii\nu_n)
\nonumber\\
&\times D_\Phi(\mathbf Q_0+\mathbf p+\mathbf K,
\ii\nu_n+\ii\Omega_m)
\nonumber\\
&\times
\Gamma_{kl}^*(\mathbf p,\ii\nu_n;\mathbf K,\ii\Omega_m).
\label{eq:Sigma_eps}
\end{align}
For acoustic polarization $e_i^\lambda(\mathbf K)$,
\begin{equation}
\mathcal P_{ij}^\lambda(\mathbf K)=
\frac12\left(K_i e_j^\lambda+K_j e_i^\lambda\right)
\label{eq:Pproj}
\end{equation}
projects the atomic displacement $u_i$ onto the strain tensor.  The mode-resolved acoustic lifetime is
\begin{equation}
\frac{1}{\tau_{\lambda\mathbf K}}
=
-\frac{1}{2\rho\Omega_{\lambda\mathbf K}}
\mathcal P_{ij}^\lambda\mathcal P_{kl}^\lambda
\Imm\,\Sigma_{\epsilon}^{ij,kl,R}(\mathbf K,\Omega_{\lambda\mathbf K}),
\label{eq:tau_def}
\end{equation}
where $\rho$ is the crystal mass density (distinct from the electron density used earlier) and $\Omega_{\lambda\mathbf K}$ is the frequency of the acoustic mode of polarization $\lambda$ at wavevector $\mathbf K$.  Eq.~\eqref{eq:tau_def} serves as the central many-body definition used in the results below.

If the vertex is approximated by its local value, $\Gamma_{ij}\simeq\lambda_0^{ij}$, the CDW part of Eq.~\eqref{eq:Sigma_eps} reduces to the composite intensity susceptibility
\begin{align}
\chi_{|\Phi|^2}(\mathbf K,\ii\Omega_m)=&
T\sum_{\nu_n}\int\frac{\dd^dp}{(2\pi)^d}
D_\Phi(\mathbf Q_0+\mathbf p,\ii\nu_n)
\nonumber\\
&\times
D_\Phi(\mathbf Q_0+\mathbf p+\mathbf K,\ii\nu_n+\ii\Omega_m).
\label{eq:chi_composite}
\end{align}
The corresponding linewidth is
\begin{equation}
\frac{1}{\tau_{\lambda\mathbf K}^{\rm loc}}
=
\mathcal A_\lambda(\mathbf K)
\left[
-\frac{\Imm\chi_{|\Phi|^2}^{R}(\mathbf K,\Omega_{\lambda\mathbf K})}
{\Omega_{\lambda\mathbf K}}
\right],
\label{eq:tau_local}
\end{equation}
where $\mathcal A_\lambda$ contains nonsingular elastic, polarization, and vertex factors.

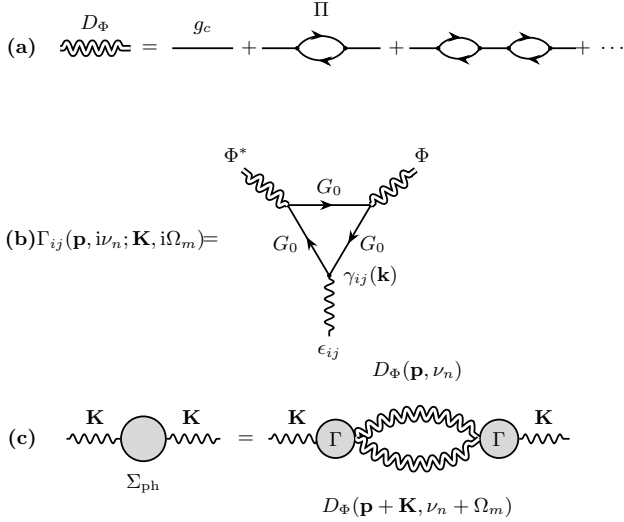
\begin{figure}[t]
\centering
\resizebox{0.98\columnwidth}{!}{%
\begin{tikzpicture}
\begin{scope}
  \node[anchor=east] at (-0.25, 0) {\textbf{(a)}};
  \draw[dphi] (0, 0) -- (1.1, 0);
  \node[above=1pt] at (0.55, 0.12) {$D_\Phi$};
  \node at (1.4, 0) {$=$};
  \draw[vc] (1.75, 0) -- (2.7, 0);
  \node[above=1pt] at (2.225, 0.08) {$g_c$};
  \node at (2.95, 0) {$+$};
  \draw[vc] (3.15, 0) -- (3.7, 0);
  \node[dot] at (3.7, 0) {};
  \draw[electron] (3.7, 0) to[out=60, in=120] (4.45, 0);
  \draw[electron] (4.45, 0) to[out=-120, in=-60] (3.7, 0);
  \node[dot] at (4.45, 0) {};
  \draw[vc] (4.45, 0) -- (5.0, 0);
  \node[above=2pt] at (4.075, 0.32) {$\Pi$};
  \node at (5.25, 0) {$+$};
  \draw[vc] (5.45, 0) -- (5.9, 0);
  \node[dot] at (5.9, 0) {};
  \draw[electron] (5.9, 0) to[out=60, in=120] (6.55, 0);
  \draw[electron] (6.55, 0) to[out=-120, in=-60] (5.9, 0);
  \node[dot] at (6.55, 0) {};
  \draw[vc] (6.55, 0) -- (7.0, 0);
  \node[dot] at (7.0, 0) {};
  \draw[electron] (7.0, 0) to[out=60, in=120] (7.65, 0);
  \draw[electron] (7.65, 0) to[out=-120, in=-60] (7.0, 0);
  \node[dot] at (7.65, 0) {};
  \draw[vc] (7.65, 0) -- (8.05, 0);
  \node at (8.45, 0) {$+\;\cdots$};
\end{scope}
\begin{scope}[yshift=-3.0cm]
  \node[anchor=east] at (-0.25, 0) {\textbf{(b)}};
  \node (glhs) at (0.95, 0) {$\Gamma_{ij}(\mathbf p,\ii\nu_n;\mathbf K,\ii\Omega_m)$};
  \node at (2.35, 0) {$=$};
  \coordinate (A) at (3.55, 0.55);
  \coordinate (B) at (4.85, 0.55);
  \coordinate (C) at (4.20, -0.55);
  \draw[electron] (A) -- (B);
  \draw[electron] (B) -- (C);
  \draw[electron] (C) -- (A);
  \node[dot] at (A) {};
  \node[dot] at (B) {};
  \node[dot] at (C) {};
  \draw[dphi] (A) -- ++(-0.7, 0.55);
  \node[above] at ($(A) + (-0.80, 0.55)$) {$\Phi^*$};
  \draw[dphi] (B) -- ++(0.7, 0.55);
  \node[above] at ($(B) + (0.80, 0.55)$) {$\Phi$};
  \draw[phonon] (C) -- ++(0, -0.95);
  \node[below] at ($(C) + (0, -0.98)$) {$\epsilon_{ij}$};
  \node[right=3pt] at (C) {$\gamma_{ij}(\mathbf k)$};
  \node[above=1pt] at ($(A)!0.5!(B)$) {$G_0$};
  \node[right=2pt] at ($(B)!0.55!(C)$) {$G_0$};
  \node[left=2pt] at ($(C)!0.45!(A)$) {$G_0$};
\end{scope}
\begin{scope}[yshift=-6.1cm]
  \node[anchor=east] at (-0.25, 0) {\textbf{(c)}};
  \node[sigmablob] (sb) at (1.3, 0) {};
  \node[below=2pt] at (sb.south) {$\Sigma_{\rm ph}$};
  \draw[phonon] (0.1, 0) -- (sb.west);
  \draw[phonon] (sb.east) -- (2.5, 0);
  \node[above=1pt] at (0.55, 0.08) {$\mathbf K$};
  \node[above=1pt] at (2.05, 0.08) {$\mathbf K$};
  \node at (2.90, 0) {$=$};
  \draw[phonon] (3.25, 0) -- (4.3, 0);
  \node[above=1pt] at (3.7, 0.08) {$\mathbf K$};
  \node[blob] (GL) at (4.3, 0) {$\Gamma$};
  \node[blob] (GR) at (6.85, 0) {$\Gamma$};
  \draw[dphi] (GL.east) to[out=40,  in=140]  (GR.west);
  \draw[dphi] (GL.east) to[out=-40, in=-140] (GR.west);
  \node[above=1pt] at ($(GL)!0.5!(GR) + (0, 0.75)$) {$D_\Phi(\mathbf p,\nu_n)$};
  \node[below=1pt] at ($(GL)!0.5!(GR) + (0, -0.75)$) {$D_\Phi(\mathbf p+\mathbf K,\nu_n+\Omega_m)$};
  \draw[phonon] (GR.east) -- (7.95, 0);
  \node[above=1pt] at (7.55, 0.08) {$\mathbf K$};
\end{scope}
\end{tikzpicture}}
\caption{Diagrammatic structure of the theory. Solid directed lines are electron Green's functions, double wavy lines are CDW fluctuation propagators, and single wavy lines are acoustic phonons. (a) The attractive charge-channel kernel generates the Gaussian CDW propagator. (b) An electron triangle couples acoustic strain to two CDW fields. (c) Two such vertices give the acoustic phonon self-energy.}
\label{fig:diagrams}
\end{figure}

\section{Results}

\subsection{Local-intensity and gradient channels}

The static channel decomposition follows from expanding the strain vertex in powers of the internal CDW-envelope momentum,
\begin{equation}
\mathcal O(\mathbf p)=\lambda_0+\lambda_1p^2+\cdots .
\label{eq:scalar_vertex}
\end{equation}
The thermal weight multiplying this vertex can be computed before making the classical approximation.  For the inertial propagator
\begin{equation}
D_\Phi(\mathbf Q_0+\mathbf p,\ii\nu_n)=
\frac{1}{\chi_\omega(\nu_n^2+E_{\mathbf p}^2)},
\qquad
E_{\mathbf p}^2=\frac{r+c p^2}{\chi_\omega},
\label{eq:inertial_sum_prop}
\end{equation}
the exact bosonic sum is
\begin{align}
\mathcal U(\mathbf p,T)\equiv
T\sum_{\nu_n}D_\Phi^2
=&\frac{1+2n_B(E_{\mathbf p})}{4\chi_\omega^2E_{\mathbf p}^3}
\nonumber\\
&+
\frac{n_B(E_{\mathbf p})[1+n_B(E_{\mathbf p})]}
{2T\chi_\omega^2E_{\mathbf p}^2}.
\label{eq:U_exact}
\end{align}
When $T\gg E_{\mathbf p}$, this reduces to
\begin{equation}
\mathcal U(\mathbf p,T)\simeq \frac{T}{(r+c p^2)^2}.
\label{eq:U_classical}
\end{equation}
Thus the commonly used static formulas are the classical, thermally populated limit of the exact Matsubara sum.

In this limit the two-vertex static weight in two dimensions is
\begin{align}
\mathcal W_{\rm stat}
&=
T\int\frac{\dd^2p}{(2\pi)^2}
\frac{(\lambda_0+\lambda_1p^2)^2}{(r+c p^2)^2}
\nonumber\\
&=\lambda_0^2I_0+2\lambda_0\lambda_1I_2+\lambda_1^2I_4.
\label{eq:Wstat}
\end{align}

The labels on these two channels reflect the operator content of the vertex rather than the integrals alone.  The $\lambda_0^2 I_0$ piece squares the constant $|\Phi|^2$ part of Eq.~\eqref{eq:SuPhi} and so couples to two insertions of the local CDW intensity; $I_0$ therefore measures how large the CDW amplitude is at a point and is the \emph{local-intensity} channel.  The $\lambda_1^2 I_4$ piece squares the gradient bilinear $\mathcal T_{ab}$ of Eq.~\eqref{eq:Gab_def}, which by $\mathcal T_{ab}=\partial_a A\,\partial_b A+A^2\partial_a\theta\,\partial_b\theta$ encodes both amplitude roughness and phase twists of $\Phi$; $I_4$ therefore measures how rapidly the CDW field varies in space and is the \emph{texture} (gradient) channel.  The cross term $2\lambda_0\lambda_1 I_2$ interpolates between the two.

Their closed forms are
\begin{align}
I_0&=\frac{T}{4\pi}\frac{\Lambda^2}{rR_\Lambda},
\label{eq:I0}\\
I_2&=\frac{T}{4\pi c^2}
\left[
\ln\left(\frac{R_\Lambda}{r}\right)+\frac{r}{R_\Lambda}-1
\right],
\label{eq:I2}\\
I_4&=\frac{T}{4\pi c^3}
\left[
c\Lambda^2+r-2r\ln\left(\frac{R_\Lambda}{r}\right)
-\frac{r^2}{R_\Lambda}
\right],
\label{eq:I4}
\end{align}
where $\Lambda$ is a circular momentum cutoff and $R_\Lambda\equiv r+c\Lambda^2$.
The long-correlation limit follows from $r=c\xi^{-2}$:
\begin{align}
I_0&\simeq \frac{T}{4\pi c r}
=\frac{T\xi^2}{4\pi c^2},
\label{eq:I0_long}\\
I_2&\simeq \frac{T}{4\pi c^2}
\ln\left(\frac{c\Lambda^2}{r}\right),
\label{eq:I2_long}\\
I_4&\simeq \frac{T\Lambda^2}{4\pi c^2}.
\label{eq:I4_long}
\end{align}
The local-intensity channel is therefore infrared enhanced, the mixed channel is logarithmic, and the gradient-gradient channel is cutoff dominated.  This hierarchy gives a simple physical interpretation: local CDW intensity fluctuations generate a narrow critical contribution, while gradient scattering is naturally tied to short-distance texture.

The opposite limit is also useful.  If $\Lambda\xi\ll1$, the denominator is nearly momentum independent across the continuum window, and
\begin{equation}
I_0\simeq \frac{T\Lambda^2}{4\pi r^2}
=\frac{T\Lambda^2\xi^4}{4\pi c^2}.
\label{eq:I0_short}
\end{equation}
At this order the integral is set by the cutoff $\Lambda$ and carries no trace of the critical scaling of Eq.~\eqref{eq:I0_long}.

The dynamic linewidth contains one additional physical ingredient: the relaxation time of the CDW fluctuation.  In the overdamped low-frequency limit, the spectral representation of Eq.~\eqref{eq:chi_composite} gives
\begin{equation}
-\frac{\Imm\chi_{|\Phi|^2}^R(0,\Omega)}{\Omega}
=
\frac12\int\frac{\dd^dp}{(2\pi)^d}
\int\frac{\dd\omega}{2\pi}
\left[-\frac{\partial n_B}{\partial\omega}\right]
\rho_\Phi^2(\mathbf p,\omega),
\label{eq:spectral_lowfreq}
\end{equation}
with $\rho_\Phi=-2\Imm D_\Phi^R$.  Using Eq.~\eqref{eq:D_overdamped} in two dimensions yields
\begin{align}
-\frac{\Imm\chi_{|\Phi|^2}^R(0,\Omega)}{\Omega}
&=
\frac{\eta T}{2}
\int\frac{\dd^2p}{(2\pi)^2}
\frac{1}{(r+c p^2)^3}
\nonumber\\
&=
\frac{\eta T}{16\pi c}
\left[
\frac{1}{r^2}
-
\frac{1}{(r+c\Lambda^2)^2}
\right].
\label{eq:dynamic_exact}
\end{align}
For $\Lambda\xi\gg1$,
\begin{equation}
-\frac{\Imm\chi_{|\Phi|^2}^R(0,\Omega)}{\Omega}
\simeq
\frac{\eta T}{16\pi c r^2}
=
\frac{\eta T\xi^4}{16\pi c^3}.
\label{eq:dynamic_xi4}
\end{equation}
The extra factor of $\xi^2$ compared with Eq.~\eqref{eq:I0_long} is the critical slowing down, $\tau_{\rm rel}\propto\xi^2$.

\subsection{Frozen texture limit and TTG scattering}

The broad TTG phenomenology is most directly connected to the gradient channel~\cite{Huang2026}.  Define the texture operator projected onto an acoustic branch as
\begin{equation}
O_{\rm tex}(\mathbf r)=
\hat n_{ij}\lambda_1^{ij;ab}
\Ree\left[(\partial_a\Phi)^*(\partial_b\Phi)\right](\mathbf r),
\label{eq:Otex}
\end{equation}
where $\hat n_{ij}$ denotes the relevant strain-polarization projection.  Its exact contribution to the linewidth has the same many-body form as Eq.~\eqref{eq:tau_local},
\begin{equation}
\frac{1}{\tau_{\lambda\mathbf K}^{\rm tex}}
=
\mathcal A_{\lambda}^{\rm tex}(\mathbf K)
\left[
-\frac{\Imm\chi_{O_{\rm tex}}^R(\mathbf K,\Omega_{\lambda\mathbf K})}
{\Omega_{\lambda\mathbf K}}
\right].
\label{eq:tau_tex_exact}
\end{equation}
The phenomenological TTG form follows when the relevant CDW texture is effectively frozen during an acoustic scattering event, so the retarded kernel is replaced by a static texture strength.

In the language of Eq.~\eqref{eq:DHO}, this frozen-texture limit applies when the soft-mode relaxation timescale set by $\Gamma_{\rm slow}$ in the overdamped regime is long compared to the acoustic period of the TTG measurement, so that the texture configuration sampled by the phonon is essentially static.  Since $\Gamma_{\rm slow}\propto r(T)$ at $\mathbf p=0$ by the mass-tracking identity Eq.~\eqref{eq:mass_tracking}, the approximation is best satisfied as the system approaches the CDW instability and progressively breaks down at high $T$ where $r(T)$ grows.

Diffraction provides that texture strength.  Let $S_\Phi(\mathbf q,T)$ be the quasielastic CDW peak with $\mathbf q$ measured relative to $\mathbf Q_0$.  Its integrated intensity is
\begin{equation}
A_{\rm diff}^2(T)\propto
\int\frac{\dd^dq}{(2\pi)^d}S_\Phi(\mathbf q,T),
\label{eq:Adiff}
\end{equation}
and its second moment is
\begin{equation}
\overline{|\nabla\Phi|^2}
=
\int\frac{\dd^dq}{(2\pi)^d}q^2S_\Phi(\mathbf q,T).
\label{eq:second_moment}
\end{equation}
If the peak shape is approximately fixed and its width is $w(T)$, then
\begin{equation}
S_\Phi(\mathbf q,T)=A_{\rm diff}^2(T)w^{-d}(T)
f\!\left(\frac{\mathbf q}{w(T)}\right),
\label{eq:peak_scaling}
\end{equation}
with $f$ normalized to unit area.

Here $A_{\rm diff}^2(T)$ sets the total peak weight and $w(T)$ its momentum width, while $f$ is a temperature-independent profile (Gaussian, Lorentzian-like, or similar); the prefactor $w^{-d}(T)$ normalizes the integrated intensity, since the substitution $\mathbf x=\mathbf q/w$ gives $\int (d^dq/(2\pi)^d)\,S_\Phi=A_{\rm diff}^2(T)\int (d^dx/(2\pi)^d)\,f(\mathbf x)$ independent of $w$.  The same rescaling applied to Eq.~\eqref{eq:second_moment} yields $q^2=w^2x^2$ and $d^dq=w^d d^dx$, so the $w^{-d}$ cancels against $w^d$ and the overall factor $w^2(T)$ of Eq.~\eqref{eq:Aw_second} emerges directly from the $q^2$ weighting of the second moment.

Equation~\eqref{eq:second_moment} becomes
\begin{equation}
\overline{|\nabla\Phi|^2}
=C_{\rm line}A_{\rm diff}^2(T)w^2(T),
\label{eq:Aw_second}
\end{equation}
where $C_{\rm line}$ depends only on the line shape.  The factor $w^2$ appears because the second moment weights the peak by the typical squared momentum inside the peak.

Combining Eqs.~\eqref{eq:tau_tex_exact} and \eqref{eq:Aw_second}, and absorbing elastic constants, line-shape factors, and scattering geometry into a prefactor, gives
\begin{equation}
\frac{1}{\tau_{\lambda\mathbf K}^{\rm tex}}
\simeq
B_{{\rm tex},\lambda}(\mathbf K)
A_{\rm diff}^2(T)w^2(T).
\label{eq:ttg_result}
\end{equation}
The phenomenological texture-scattering rate of the TTG analysis~\cite{Huang2026} follows from the local limit of Eq.~\eqref{eq:ttg_result}.  It is distinct from the local-intensity channel, so a minimal two-channel parameterization is
\begin{equation}
\tau_{\rm CDW}^{-1}(T)
\simeq
B_{\rm tex}A_{\rm diff}^2(T)w^2(T)
+
B_0\mathcal X_{\rm loc}(T),
\label{eq:two_channel_fit}
\end{equation}
where $\mathcal X_{\rm loc}$ is determined by the local composite response.  In the quasistatic Gaussian limit $\mathcal X_{\rm loc}\propto T/r$, while in the overdamped low-frequency limit $\mathcal X_{\rm loc}\propto T/r^2$.

\subsection{IXS soft-mode observables}

IXS near the CDW wavevector probes the one-particle lattice spectral function, not the TTG acoustic lifetime~\cite{Weber2023}.  The measured intensity is proportional to
\begin{equation}
S_{uu}(\mathbf Q_0+\mathbf p,\omega)
\propto
[1+n_B(\omega)]
\left[-\Imm G_{uu}^R(\mathbf Q_0+\mathbf p,\omega)\right].
\label{eq:IXS}
\end{equation}
The many-body information enters through the phonon self-energy, which in the CDW channel takes the schematic form $\Sigma_{uu}^R(\mathbf q,\omega)\simeq g_{\rho u}^2\,\chi_\rho^R(\mathbf q,\omega)$; equivalently, $G_{uu}^R$ and the CDW response share the determinant of the coupled $u$--$\rho$ matrix in Eq.~\eqref{eq:shared_poles}.  Near the soft point $(\mathbf Q_0,0)$, expanding that inverse propagator to leading order in $\mathbf p=\mathbf q-\mathbf Q_0$ and $\omega$ gives
\begin{equation}
[G_{uu}^R]^{-1}(\mathbf Q_0+\mathbf p,\omega)\;\propto\; r(T)+c\,\mathbf p^2-\chi_\omega\,\omega^2-\ii\eta\,\omega,
\label{eq:Guu_expanded}
\end{equation}
where $c$ is the envelope stiffness, $\chi_\omega$ the inertial coefficient, and $\eta$ the overdamped damping coefficient, all of which are slowly varying in this window.  The mass $r(T)=g_c^{-1}-\Pi(\mathbf Q_0,0;T)$ is precisely that of Eq.~\eqref{eq:rdef}: it is generated by the many-body susceptibility, so as temperature changes $\Pi(\mathbf Q_0,0;T)$ changes, $r(T)$ shifts, and the pole of the phonon spectral function moves with it.  The temperature evolution seen by IXS is therefore not an extra phenomenological input but an inheritance from the CDW susceptibility through $r(T)$.

Dividing Eq.~\eqref{eq:Guu_expanded} by the smooth inertial prefactor $\chi_\omega$ puts the denominator in the damped-harmonic-oscillator form of Eq.~\eqref{eq:DHO}, $\widetilde\omega_q^2-\omega^2-2\ii\Gamma_q\omega$, with the identifications
\begin{equation}
\widetilde\omega_q^2=\frac{r(T)+c\,\mathbf p^2}{\chi_\omega},\qquad 2\Gamma_q=\frac{\eta}{\chi_\omega},
\label{eq:DHO_ident}
\end{equation}
with $\widetilde\omega_q$ the natural frequency and $\Gamma_q$ the damping of the soft CDW--lattice mode of Eq.~\eqref{eq:DHO}, so the IXS peak tracks the hybrid CDW--lattice soft pole.  In the underdamped fitting window, Eq.~\eqref{eq:DHO_ident} at $\mathbf p=0$ gives
\begin{equation}
\widetilde\omega_q^2(\mathbf Q_0,T)=\frac{r(T)}{\chi_\omega}.
\label{eq:IXS_omega}
\end{equation}
If the mass is locally well approximated by $r(T)=a(T-T_0)$, then the IXS soft-mode energy squared is linear in temperature within that window.  The intercept $T_0$ is a fitting scale, not necessarily the experimentally named full-softening temperature~\cite{Weber2023}.  If the same branch is overdamped, the slow pole in the damped-oscillator convention of Eq.~\eqref{eq:DHO} is
\begin{equation}
\Gamma_{\rm slow}
=
\Gamma_q-\sqrt{\Gamma_q^2-\widetilde\omega_q^2}
\simeq
\frac{\widetilde\omega_q^2}{2\Gamma_q},
\qquad
\widetilde\omega_q\ll\Gamma_q.
\label{eq:slow_pole}
\end{equation}
Physically, $\Gamma_{\rm slow}$ is the slow quasielastic relaxation rate of the overdamped CDW amplitude: when $\widetilde\omega_q<\Gamma_q$ the two roots of Eq.~\eqref{eq:DHO} collapse onto the imaginary axis, and $\Gamma_{\rm slow}$ governs the exponential decay of CDW amplitude fluctuations toward equilibrium, with $\Gamma_{\rm slow}\to 0$ as $r(T)\to 0$ since $\widetilde\omega_q^2\propto r$ vanishes at the would-be transition.  The direct mass-tracking quantity in the overdamped regime is therefore the product
\begin{equation}
2\Gamma_q\,\Gamma_{\rm slow}=\widetilde\omega_q^2=\frac{r(T)}{\chi_\omega},
\label{eq:mass_tracking}
\end{equation}
while $\Gamma_{\rm slow}$ alone inherits the $r(T)$ dependence only to the extent that $\Gamma_q(T)$ varies slowly across the fitting window.  Thus IXS tests whether $\widetilde\omega_q^2$, or the product $2\Gamma_q\Gamma_{\rm slow}$ in the overdamped case, follows the same local mass $r(T)$.  The relation applies to the soft branch near $\mathbf Q_0$ measured by IXS~\cite{Weber2023}, and does not extend to the full high-temperature residual diffuse CDW regime reported in thermal transport and diffraction~\cite{Huang2026}.

\section{Discussion}

The microscopic content of the theory is summarized by three structural equations.  The CDW fluctuation propagator $D_\Phi^{-1}=V_c^{-1}-\Pi$ of Eq.~\eqref{eq:Dphi_inv} encodes the CDW instability through the peak of $\Pi$ (or of $V_c\Pi$ when the electron--phonon coupling is itself momentum-selective) at $\mathbf Q_0$; the small parameter $r(T)$ of Eq.~\eqref{eq:rdef} measures how close the system is to that instability and sets the correlation length of Eq.~\eqref{eq:xi_r}.  The strain--intensity vertex $S_{u\Phi}$ of Eq.~\eqref{eq:SuPhi}, with its gradient decomposition Eq.~\eqref{eq:grad_decomp}, follows from the electron triangle $\Gamma_{ij}$ of Eq.~\eqref{eq:Gamma} by long-wavelength expansion and translational invariance; it is the microscopic origin of any local elastic-coupling functional $\delta C_{ijkl}(\mathbf r)$ whose coefficients vary with the CDW amplitude.  The phonon self-energy $\Sigma^{ij,kl,R}_\epsilon$ of Eq.~\eqref{eq:Sigma_eps} is the susceptibility of the composite CDW intensity, Eq.~\eqref{eq:chi_composite}, weighted by the strain-coupling kernel; through Eq.~\eqref{eq:tau_def} it sets the mode-resolved acoustic lifetime.  Together these equations tie the CDW instability, the spatial inhomogeneity of the CDW amplitude, and the phonon damping to a single microscopic origin.

The framework gives one common language for three measurements that are usually discussed in isolation.  Inelastic X-ray scattering near the CDW wavevector resolves the lattice projection of the hybrid one-particle soft pole through the damped-harmonic-oscillator structure of Eqs.~\eqref{eq:IXS}--\eqref{eq:IXS_omega}, with the underdamped--to--overdamped crossover and the mass-tracking product Eqs.~\eqref{eq:slow_pole}--\eqref{eq:mass_tracking} as direct experimental signatures of the same $r(T)$.  Transient thermal grating measures the long-wavelength attenuation of acoustic phonons through the local two-channel fit of Eq.~\eqref{eq:two_channel_fit} and the TTG closed form of Eq.~\eqref{eq:ttg_result}, dominated by the UV-controlled gradient channel $I_4$ of Eq.~\eqref{eq:I4_long} but also containing a critical contribution from the local channel $I_0$ of Eq.~\eqref{eq:I0_long}.  Diffuse X-ray scattering measures the static structure factor of the CDW amplitude, whose second moment Eq.~\eqref{eq:Aw_second} fixes the phenomenological $A^2(T)w^2(T)$ background of the TTG analysis~\cite{Huang2026} in terms of the same $r(T)$ and $c$.  All three observables are governed by the same propagator $D_\Phi$ and the same strain coupling, so quantitative consistency among them is a sharp test of the picture.

The construction is structurally parallel to paramagnon-mediated transport in nearly ferromagnetic metals~\cite{BealMonod1968,FayLayzer1968,DoniachEngelsberg,BerkSchrieffer,Moriya} but differs from it in three physically essential respects, each rooted in a feature of $\Pi(\mathbf q)$ or of the strain vertex.  First, the CDW precursor carries a finite ordering wavevector $\mathbf Q_0$, so that phonons scatter through large-momentum-transfer processes that relax the heat current with backscattering weight of order unity, with no Ward-identity cancellations of the kind that protect conserved currents in the long-wavelength magnetic case.  Second, the relaxation mechanism is not an internal-index flip but the modulation of the local elastic environment by the spatially inhomogeneous CDW intensity entering Eq.~\eqref{eq:chi_composite}.  Third, translational invariance forces the acoustic coupling to act through strain rather than through displacement, so the self-energy of Eq.~\eqref{eq:Sigma_eps} responds only to spatial variations of the CDW amplitude and is blind to its mean value.  These features generalize unchanged to any peaked-$\Pi$ instability in a correlated metal, so the same diagrammatic skeleton applies to the rare-earth tritellurides~\cite{Brouet2008}, kagome CDW compounds~\cite{Ortiz2020}, and the fluctuating charge-order regime of cuprates~\cite{Ghiringhelli2012,Chang2012}, with the material entering only through the bandstructure that sets $\Pi$ and through the elastic moduli that contract with $\mathcal P^\lambda_{ij}$.

The picture differs qualitatively from strong-coupling Ising-type scenarios for fluctuating CDW regimes, most notably Gor'kov's two-minimum local-potential model~\cite{Gorkov2012} in which the transition is an Ising ordering of discrete ionic states.  In that scenario there is no soft collective electronic mode and no underdamped--overdamped crossover; the present framework predicts both, with $\tilde\omega_q^2$ tracking $r(T)$ in the underdamped window of Eq.~\eqref{eq:IXS_omega} and migrating, when $\Gamma$ becomes comparable to $\tilde\omega_q$, into a slow quasielastic rate $\Gamma_{\rm slow}\simeq\tilde\omega_q^2/(2\Gamma)$ of Eq.~\eqref{eq:slow_pole} that obeys the mass-tracking identity Eq.~\eqref{eq:mass_tracking} by construction.  The IXS data of Ref.~\cite{Weber2023} on 2H-TaSe$_2$ display the underdamped linear-in-$T$ softening of the squared mode energy; the present picture predicts that DHO fits separating $\tilde\omega_q$ from $\Gamma$ should reveal $\Gamma_{\rm slow}(T)$ continuing linearly toward zero at the same intercept $T^*$ once the mode crosses into the overdamped regime.  This is a sharp test that the Ising scenario does not pass, and which can be performed in any CDW metal for which IXS soft-mode spectroscopy is available.

The recent TTG measurements on 2H-TaSe$_2$ of Ref.~\cite{Huang2026} are recovered by the framework and supplemented by it.  The phenomenological $\delta C_{ijkl}(\mathbf r)\,\epsilon_{ij}\epsilon_{kl}$ coupling between acoustic strain and the spatially fluctuating CDW amplitude introduced there is precisely the field-theory structure of Eq.~\eqref{eq:SuPhi}, with the amplitude--gradient decomposition of Eq.~\eqref{eq:grad_decomp} mapping onto the texture parameters used in the TTG analysis.  The two-channel fit of Eq.~\eqref{eq:two_channel_fit} sharpens the connection: the broad UV-dominated background that the $A^2(T)w^2(T)$ phenomenology approximates is the gradient channel $I_4$ of Eq.~\eqref{eq:I4_long}, set by the cutoff $\Lambda$ in the long-coherence-length limit $\xi\Lambda\gg 1$ where Eq.~\eqref{eq:I4_long} reduces to $I_4\simeq T\Lambda^2/(4\pi c^2)$ and the contribution is only weakly $T$-dependent, acting as a broad cutoff-dominated background appropriate to the long-range-ordered regime in which $\xi$ is large; this same channel acquires an additional temperature signature in the opposite short-coherence-length regime $\xi\Lambda\sim 1$ characteristic of the fluctuation precursor, where the subleading $r$-dependent and $\ln(R_\Lambda/r)$-dependent pieces of Eq.~\eqref{eq:I4} are no longer negligible and the background itself develops temperature structure beyond the linear prefactor, while the narrow critical channel $I_0$ of Eq.~\eqref{eq:I0_long} predicts an additional sharp excess scattering $\propto T/r(T)\propto T\xi^2(T)$ peaked near the precursor regime.  The latter is genuinely new and provides a direct experimental test: subtracting the diffraction-fixed background from the TTG diffusivity should reveal the narrow critical residual, with its temperature profile a direct image of $r(T)$.  The microscopic theory therefore not only supplies the origin of the elastic coupling that Ref.~\cite{Huang2026} treated as a fit parameter, but also sharpens its predictive content by separating the broad and the critical contributions.

What the Gaussian theory does not supply in closed form is a first-principles prediction of $r(T)$ across the full temperature window outside the linear softening regime.  The full nonmonotonic $\kappa(T)$ reported for 2H-TaSe$_2$~\cite{Huang2026}, including the recovery at high temperature, is reproduced through the gradient channel once the measured $\xi(T)$ is supplied as input, but a microscopic determination of $\xi(T)$ requires extensions beyond the present skeleton.  Three are natural.  A self-consistent treatment of the quartic invariant in the Landau--Ginzburg expansion of $D_\Phi$ would renormalize $r(T)$ near the transition and round the mean-field divergence into a first-order jump; an interlayer stiffness term in $D_\Phi^{-1}$ would capture the dimensional crossover that softens the narrow anomaly at higher temperature, with the qualitative consequence that the two-dimensional kinematics responsible for the strong $T\xi^2$ enhancement of $I_0$ would weaken parametrically as the system becomes more three-dimensional, predicting a stronger critical anomaly in thinner samples or in the presence of weaker interlayer coupling; quenched disorder $\delta r(\mathbf r)$ would generate an additional smooth background and round the cusp at the first-order transition.  The framework itself is not tied to 2H-TaSe$_2$: it natively covers any CDW instability driven by a peaked electronic susceptibility, including both nesting and saddle-point scenarios as catalogued in Refs.~\cite{Rossnagel2011,JohannesMazin2008}, and generalizes immediately to momentum-selective electron--phonon coupling.  Application to the other transition-metal dichalcogenides~\cite{MoncAxeDiSalvo1977,Monceau2012}, the rare-earth tritellurides~\cite{Brouet2008}, the kagome CDW compounds~\cite{Ortiz2020}, and the cuprate fluctuating charge-order regime~\cite{Ghiringhelli2012,Chang2012} should follow the same electron-loop construction that governs the interplay between the CDW instability and the heat-carrying phonons, with high-resolution resonant inelastic x-ray scattering (RIXS) on the cuprates and beyond~\cite{Arpaia2019,Fedele2026} now providing direct momentum- and energy-resolved access to the soft CDW--lattice mode and thereby a stringent experimental test of the hybrid soft-mode propagator and strain--intensity vertex developed here.

\begin{acknowledgments}
This work was motivated by ongoing efforts to understand the IXS observations of acoustic phonon softening and the anomalous thermal transport in 2H-TaSe$_2$ at elevated temperatures.  The author thanks Jinghang Dai and Zhiting Tian for useful discussions on experimental data.
\end{acknowledgments}

\bibliography{references}

\end{document}